%% file: ON_OFF_Privacy_with_Correlated_Requests.tex
\documentclass[conference]{IEEEtran}
\input{preamble}

\begin{document}

\title{ON-OFF Privacy with Correlated Requests} 


 \author{
   \IEEEauthorblockN{Carolina Naim, Fangwei Ye, Salim El Rouayheb}
   \IEEEauthorblockA{Department of Electrical and Computer Engineering, Rutgers University\\
                     Emails: \{carolina.naim, fangwei.ye, salim.elrouayheb\}@rutgers.edu
                     }
                    
}

\maketitle

\begin{abstract}

 We introduce the ON-OFF privacy problem. At each time, the   user is interested in  the latest message of one of $N$ online sources chosen at random, and his privacy status can be ON or OFF  for each request. Only when privacy is ON the user wants to hide the source he is interested in. The problem is to design ON-OFF privacy schemes with maximum download rate  that allow the user to obtain privately his requested messages. In many  realistic scenarios,  the user's requests are correlated since they depend on his personal attributes such as age, gender, political views, or geographical location. Hence,  even when privacy is OFF,  he  cannot simply reveal his request since this will leak information about his requests when privacy was ON. We study the  case when the users's requests can be modeled by a Markov chain and  $N=2$ sources. In this case, we propose an ON-OFF privacy scheme and prove its optimality.

%
\end{abstract}

%

\section{Introduction}
\label{sec:introduction}

\subsection{Motivation}

Privacy is a major concern for online users who can unknowingly reveal critical personal information (age, sex, diseases, political proclivity, etc.) through daily online activities such as watching online videos, following people and liking posts on social media, reading news and searching websites. This is now a well-acknowledged concern and has lead to many interesting theoretical problems such as anonymity \cite{Sweeney_2002}, differential privacy \cite{Dwork_2006}, private information retrieval \cite{Chor_1995}, and other privacy-preserving algorithms.

In all these formulations the user is assumed to always want to maintain a certain level of privacy, which we refer to as privacy being always ON. However, in many scenarios, the user may wish to switch between privacy being ON and OFF. This switch depends on several criteria such as location, network/connection or phone/machine being used, to name a few. The reason the user may want to flip between these two modes, instead of keeping privacy always ON, is that typically privacy-preserving solutions incur a degradation in the quality of service, mostly felt by the user through large delays. Service providers may also be interested in incentivizing  the user to require privacy only when it is needed since private solutions also incur higher communication and computation costs on their side.

One may be tempted to propose the simple solution in which the user has available to him two schemes, one private and one non-private. Over time, the user simply switches between these two schemes depending on whether privacy is turned ON or OFF. The problem with this solution is that it  guarantees privacy only if the user's online activities are statistically independent over time. However, a user's online activities are typically personal, making them correlated over time. For example, a bilingual English/Spanish user, who is  checking  the news in Spanish now, is more likely to keep reading the news in Spanish for a while before switching to English. At that point English becomes more probable.  Another example is when  the user   is watching online videos. The user chooses the  video to watch next from a  list of videos recommended to him  and this list depends on previously watched videos.  Thus,  due to correlation, simply ignoring the privacy requirement when privacy is OFF may reveal information about the activities when privacy was ON.

\subsection{Example}
\label{sec:example}
To be more concrete and to  gently introduce our setup for ON-OFF privacy, we give the following example.  Suppose a user is watching political or news videos online. At each time $t$, the user has a choice between two new videos each of which is produced by two different news sources, $A$ and $B$.  Source $A$  is politically left-leaning and source $B$ is right-leaning.  

\begin{figure}[b]
\vspace{-1pt}
\centering
\begin{tikzpicture}[thick,scale=0.8, every node/.style={scale=0.8}]
 \tikzstyle{every node}=[font=\normalsize]
        \tikzset{node style/.style={state, 
        							inner sep=0.5pt,
                                    minimum size = 25pt,
                                    line width=0.2mm,
                                    fill=white},
                    LabelStyle/.style = { minimum width = 1em, fill = white!10,
                                            text = black},
                   EdgeStyle/.append style = {->, bend left=22, line width=0.2mm} }

        \node[node style] at (4,7)     (pro-left)     {A};
        \node[node style] at (8, 7)     (pro-right)     {B};
		  \Edge[label = $\alpha$](pro-left)(pro-right)
  			\Edge[label = $\beta$](pro-right)(pro-left)
            
            \Loop[dist = 2.2cm, dir = NO, label = $1-\alpha$](pro-left.west)
            \Loop[dist = 2.2cm, dir = SO, label = $1-\beta$](pro-right.east)           
    \end{tikzpicture}
    \caption{The two-state Markov chain representing the  correlation of the user's requests $X_t, t \in \bbZ$.  }
\label{fig:example}
\end{figure}
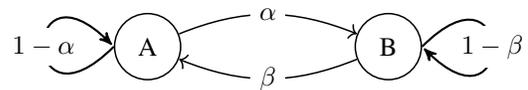

Let $X_t\in\{A,B\}$ be the source whose video the user wants to watch at time $t \in \bbZ$. We model the correlation among the user's requests by assuming that $X_t$ is the two-state Markov chain  depicted in Figure~\ref{fig:example}, where $\alpha=\Pr(X_{t+1}=B\mid X_t=A)$ and 
$\beta=\Pr(X_{t+1}=A\mid X_t=B)$.
For example, we choose $\alpha=\beta=0.2$. This means that if the current video being watched is left-leaning, there is an $80\%$ chance that the next video is also left-leaning, and vice versa. 

For the sake of brevity, we focus on the two time instants $t=0$ and $t=1$, and assume that privacy is  ON at $t=0$ and is switched to OFF at $t=1$. This means that the user would like to hide whether he was watching a left-leaning or a right-leaning video at time $t=0$, but does not care about revealing the source of the  video he watched at $t=1$.

  The goal is to devise an ON-OFF privacy scheme that  always gives the user the video he wants, but never reveals the choice of sources when privacy is ON, \Ie $t=0$ in this case. We are interested in schemes that minimize the download cost, or equivalently maximize the download rate (the inverse of the normalized download cost). 

%

At  $t=0$, the problem is simple. The user  achieves privacy by downloading both videos. We say that the user's  query at $t=0$ is $Q_{0}=AB$. Therefore, the download rate at $t=0$ is $R_0=1/2$.

At  $t=1$, the privacy is OFF. Now, the user must be careful not to directly declare his request, because this may reveal information about his  request at $t=0$ which is to remain private. The user can again download both videos, \Ie $Q_{1}=AB$,  and achieve privacy with a rate $R_1=1/2$. 

Our key result is that the user can achieve a better rate at $t=1$, without compromising privacy,  by 
\begin{itemize}
\item choosing randomly between downloading  $A$ ($Q_{1}=A$) or both $A$ and $B$ ($Q_{1}=AB$) if he wants $X_1=A$,
\item choosing randomly between downloading  $B$ ($Q_{1}=B$) or both $A$ and $B$ ($Q_{1}=AB$) if he wants  $X_1=B$. 
\end{itemize}

%
%
%
%
%
\begin{table}[t]
\normalsize
\centering
	\begin{tabular}{ c c |c c c }
	$X_{0}$ & $X_1$& $Q_1=A$ &$Q_1=B$& $Q_1=AB$\\
	 \toprule
	 $A$ & $A$ &$0.25$&$0$&$0.75$\\
	 $A$ & $B$& $0$ &$1$ &$0$\\
	 $B$ & $A$&$1$&$0$&$0$\\
	 $B$ & $B$&$0$&$0.25$&$0.75$\\

	\end{tabular}
	\caption{An example of our ON-OFF privacy scheme for $\alpha=
\beta=0.2$. The query $Q_{1}$ at $t=1$ is a probabilistic function of $X_0$ and $X_1$, the requests at $t=0$ and $t=1$ respectively. The entries of the table represent the probabilities $p(Q_{1}\mid X_{0},X_1)$. $Q_1=AB$ means that the user downloads the videos from both sources $A$ and $B$. }
	\label{table:example}
	\vspace{-10pt}
\end{table}

This random choice must also depend on the request $X_0$ at $t=0$.  The different probabilities defining the    scheme are given  in Table~\ref{table:example} and will be justified later when we explain the general scheme. For now, one can check that these probabilities lead to 
$$\Pr(Q_1 =q)=\Pr(Q_1 =q\mid X_0 =x_0),$$
 for  any $q \in \{A, B, AB\}$ and any $x_0 \in \{A, B\}$. Thus, $X_0$ and $Q_1$ are independent and  the proposed scheme in  Table~\ref{table:example} achieves perfect privacy for the request at $t=0$. 
 Moreover, the scheme ensures that the user  always obtains  the video he is requesting.





For $t=1$, the rate  $R_1=1/(2-\alpha-\beta)=0.625$, which is strictly greater than $0.5$, the rate of querying  both files. We later show that this rate is actually optimal. In fact, the values in Table \ref{table:example} were carefully chosen to achieve the privacy at the highest download rate. Any other choice of the probabilities $p(Q_{1}\mid X_{0},X_1)$ would either violate privacy or lose the optimality of the rate.


\subsection{Setup \& Contributions}

We introduce a mathematical model to capture the ON-OFF privacy problem when the user is downloading data from online sources.

We consider the setup in which there are $N$ information sources each producing a new message at each time $t \in \bbZ$. At each time $t$, the user randomly chooses one of the sources and requests its latest produced message.

The privacy constraint is the following: the user wants to leak zero information about the identity of the sources in which he is interested  at each time $t$ when the privacy is ON. The main challenge stems from the fact that the user's requests are not independent. As in the previous example, we model the  dependence between these requests by an $N$-~state Markov chain. The goal is to design an ON-OFF privacy scheme with maximum download rate that satisfies the user's request and guarantees the privacy of the requests made when privacy is ON.

Our technical results can be summarized as follows. We study the case of $N=2$ sources for the special but important case where privacy is ON for $t\leq0$ and switched OFF for $t\geq1$. We prove an upper bound on the instantaneous download rate at each time $t$, and give an ON-OFF privacy scheme that achieves it.


\subsection{Related Work}
 The special case of the ON-OFF privacy problem in which privacy is always ON and the user's requests are independent  reduces to the  information-theoretic private information retrieval (PIR) problem  on a single server. In this case, the user cannot do anything smarter than downloading everything \cite{Chor_1995} (except  the recently studied problem when the user has side information \cite{Kadhe_2017} which is not the case here). Recently, there has been significant research activity on determining the maximum download rate of PIR with  multiple servers  (e.g. \cite{Shah_2014, Sun_2017, Tajeddine_2016, Freij-Hollanti_2017, Banawan_2018}). However,  the model there requires multiple servers and,  in the parlance of this paper,  privacy is assumed to be  always ON.

\section{Problem Formulation and notations}

\label{sec:formulation}

The ON-OFF privacy model can be described as follows. A single server stores $N$ sources indexed by $\cN:=\{1,\ldots,N\}$. Each source generates a message $W_{x,t}$ at time $t$, where $x~ \in~ \cN$. We only consider a discrete time throughout this paper, \Ie $t \in \bbZ$. For any integers $a$ and $b$ such that $a \leq b$, denote $\{a,\ldots,b\}$ by $[a:b]$, and $\{i: i= a,a-1,a-2,\dots\}$ by $(a)$.

A user retrieves messages consecutively from the server. He is interested in one of the sources at each time, and wishes to retrieve the latest message generated by the  desired source.

In particular, let $X_t$ be the index of the desired source at time $t$, which takes values in $\mathcal{N}$, and in the sequel we call $X_t$ the user's request. By slightly abusing the notation, we denote the latest message generated by the desired source $X_t$ as $W_{X_t,t}$, and the user wishes to retrieve the message $W_{X_t,t}$. We assume that the messages $\{W_{x,t}: x \in \cN, t \in \bbZ\}$ are mutually independent, each of which consists of $L$ symbols. Without loss of generality, we assume that each of the messages is uniformly distributed on $\{0,1\}^L$, \Ie
\begin{equation}
 	H\left(W_{x,t}: x \in \cN, t \in \bbZ \right) = \sum_{x,t} H\left(W_{x,t}\right),
 \end{equation}
  \vspace{-15pt}

 \hspace{-10pt}and
 
  \vspace{-10pt}
\begin{equation}
	H\left(W_{x,t}\right) = L.
\end{equation}

As discussed in Section~\ref{sec:introduction}, we are particularly interested in the case where the requests $X_t$, for $t \in \bbZ$, form a Markov chain. The transition matrix of the Markov chain is known to both the server and the user.  

Meanwhile, the user may or may not wish to keep the identity of the source he is interested in at time $t$, hidden from the server. Specifically, the privacy mode $F_t$ at time $t$ can be either ON or OFF, where $F_t$ is ON when the user wishes to keep $X_t$ private, while $F_t$ is OFF when the user is not concerned with privacy.

In this paper, we focus on  the case in which the privacy mode is the step function given by .
\begin{equation}
\label{eq:step-function}
F_t =
\begin{cases}
\text{ON},  &t \leq 0,\\
\text{OFF},  &t \geq 1.
\end{cases}
\end{equation} 
Solving the problem for this step function is an essential building block for tackling the general case where $\{F_t: t \in \bbZ\}$ is a random process. A discussion about the general case can be found 
in  Appendix \ref{app:discussion}.
 
The user is allowed to generate unlimited local randomness and we are not interested in the amount of randomness used. Therefore, we assume without loss of generality that the random variables $\{S_t : t \in \bbZ\}$, representing the local randomness, are mutually independent. Moreover, we assume that the user's requests $\{X_t:t \in \bbZ\}$, the messages $\{W_{x,t}: x \in \cN, t \in \bbZ\}$ and the local randomness $\{S_t: t \in \bbZ\}$ are mutually independent. 


As discussed in Section~\ref{sec:introduction}, if the user carelessly downloads the desired message at time $t$ when the privacy is OFF, the privacy in the past may be compromised.
To ensure privacy, the user may utilize the requests $\{X_i: i \leq t \}$ and the local randomness $\{S_i: i \leq t \}$ to construct a query $Q_t$ and send it to the server. Upon receiving the query, the server responds to the request by producing the answer $Y_t$ consisting of $\ell\left(Q_t\right)$ symbols, where the length of $Y_t$ is a function of the query $Q_t$. Thus, the average length of the answer $Y_t$ is given by
\begin{equation}
  \ell_{t} = \mathbb{E}_{Q_{t}} [\ell\left(Q_t\right)].
\end{equation}

The query $Q_t$ at time $t$ is assumed to be a function of all the requests $\{X_i: i \leq t \}$ and all the local randomness $\{S_i: i \leq t \}$ up to and including time $t$, \Ie
\begin{equation}
	Q_t = \phi_t \left(X_{(t)},S_{(t)} \right).
\end{equation}
Note that since the previous answers $\{Y_i: i < t \}$ are functions of the previous messages, which are independent on the current message, the previous answers will not help in retrieving the current message, so without loss of generality, $Q_t$ is not encoded from $\{Y_i: i < t \}$.



Correspondingly, the answer $Y_t$ of the server is a function of the query $Q_t$ and the messages $\{W_{x,t}: x \in \cN\}$, \Ie 	
\begin{equation}
 	Y_t = \rho_t \left(Q_t,W_{1,t},\ldots, W_{N,t}  \right).
\end{equation} 

These functions need to satisfy the decodability and the privacy constraints, \Ie
\begin{enumerate}
	\item Decodability: For any time $t$, the user should be able to recover the desired message from the answer with zero-error probability, \Ie
	\begin{equation}
	\label{eq:decode}
		H\left(W_{X_t,t}|Y_{t}\right) = 0, \quad \forall t\in \bbZ.
	\end{equation}
	\item Privacy: For any time $t$, given all past queries received by the server, the query $Q_t$ should not reveal any information about all the past or present requests where the privacy is ON, that is
	\begin{equation}
	\label{eq:privacy}
		I \left(X_{\cB_t};Q_t|Q_{(t-1)}\right) = 0, \quad \forall t\in \bbZ,
	\end{equation}
	where $\cB_t = \{i:i \leq t,F_i=\text{ON}\}$.
\end{enumerate}

For any message length $L$, the tuple $\left(\ell_t: t \in \bbZ\right)$ is said to be achievable if there exists a code satisfying the decodability and the privacy constraint. 
The efficiency of the code can be measured by the download rate $R_t:=\frac{L}{\ell_t}$. Hence, we define the achievable region as follows:

\begin{definition}
The rate tuple $\left(R_t:t \in \bbZ\right)$ is achievable if there exists a code with  message length $L$ and average  download cost $\ell_t$ such that $R_t \leq \frac{L}{\ell_t}$. 
\end{definition}
Conventionally, the capacity region $\sC \left( \cP \right)$ can be defined as the closure of the set of achievable rate tuples $\left(R_t:t \in \bbZ\right)$, where $\cP$ is the set of all possible probability distributions of $p\left(X_t: t \in \bbZ \right)$.  Table \ref{table:nomenclature} summarizes our notation.


\begin{table}[t]
\centering
\small

\begin{tabular}{ l|l}

Symbol & Definition\\
\toprule
 $N$ & Number of sources \\[0.5ex]

$W_{x,t}$ & Message generated by source  of index $x$ at time $t$\\[0.5ex]

$X_t$ & User's request at time $t$ ($X_t \in \cN $)\\[0.5ex]

$F_t$ & Privacy mode at time $t$ (ON or OFF)\\[0.5ex]

 $Q_t$ & Query sent by the user to the server at time $t$\\[0.5ex]
 
 $Y_t$ & Answer sent by the server to the user at time $t$\\[0.5ex]
 
 $S_t$ & Local randomness generated by the user at time $t$\\[0.5ex]
 
 $\ell_{t}$ & Average length of the answer $Y_t$\\[0.5ex]
 
 $R_t$ & Rate at time $t$\\[0.5ex]
 
 $[a:b]$ & $=\{a,\ldots,b\}$ for any integers $a$ and $b$ such that $a \leq b$\\[0.5ex]
 
 $(a)$ & $=\{i: i= a,a-1,a-2,\dots\}$ for any integer $a$\\[0.5ex]
\toprule

\end{tabular}
\caption{Nomenclature and Notation}
\label{table:nomenclature}
\end{table}

\section{Main results }

Our main result is a complete characterization of the achievable region for the case of two sources, \Ie $N=2$.
We will use $A$ and $B$ to denote these two sources. In this case, the requests $X_t$ follow a two state Markov chain defined by the transition matrix 
\begin{equation}
\label{eq:Source-Markov}
 M =
\begin{bmatrix}
       1- \alpha &  \alpha         \\
       \beta  & 1- \beta 
\end{bmatrix},
\end{equation}   
where $\alpha$ is the transition probability from $A$ to $B$, and $\beta$ is the transition probability from $B$ to $A$.

We first state the main theorem of this paper.
\begin{theorem}
\label{thm:main}
	For privacy mode given in \eqref{eq:step-function}, the rate tuple $\left(R_t:t \in \bbZ\right)$ is achievable if and only if  
  \begin{equation}
      R_t \leq 
      \begin{cases}
      \frac{1}{2}, & t \leq 0 ,\\
        \frac{1}{1+ |1- \alpha -\beta|^t}, & t \geq 1.
      \end{cases}
  \end{equation}
\end{theorem}

\begin{figure}[!t]
    \centering
    \input{plot.tex}
    \caption{The maximum rate $R_t$, as given in Theorem \ref{thm:main}, as a function of time for different values of $\alpha+\beta$. As $\alpha+\beta$ approaches $1$, the correlation between the request decreases leading to an increase in the rate. For $\alpha+\beta=1$ the requests are independent. In this case, when privacy is ON at $t=0$, the user downloads messages from both sources ($R_t=1/2$), and for $t>0$, privacy is OFF and the user downloads only the message he wants ($R_t=1$).}
    \label{fig:th_plot}
    \vspace{-10pt}
\end{figure}
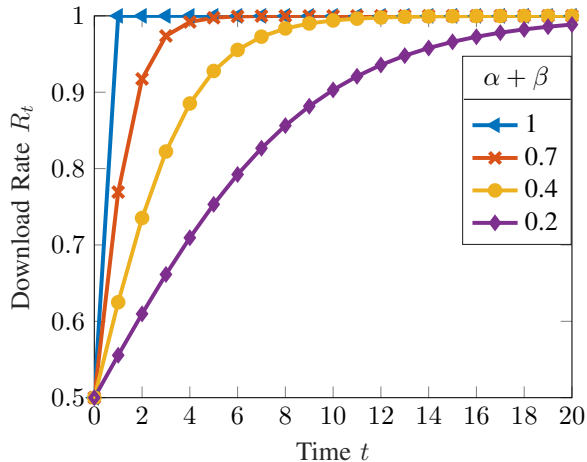
When privacy is ON, for $t\leq 0$, the user has to request both the most recent messages of $A$ and $B$. Therefore, the rate $R_t=1/2$. 

The more interesting part of Theorem \ref{thm:main} is for $t\geq1$. For a fixed time $t\geq1$, the rate as a function of $\alpha$ and $\beta$ is symmetric around $\alpha+\beta=1$. When $\alpha+\beta=1$, the user's requests are independent such that $p(X_t\mid X_{t-1})=p(X_t)$, so the user can directly query for his desired message, \Ie $Q_t=X_t$. The rate is then maximized to $R_t=1$. 

In terms of asymptotics,  when the Markov chain is ergodic, the download rate goes to $1$ as $t$ goes to infinity. Intuitively, as $t$ grows, the information carried by $X_t$ about $X_0$ decreases, so the user can eventually directly query for what he wants, \Ie $Q_t=X_t$. Otherwise, when the Markov chain is not ergodic ($\alpha=\beta=0$ or $\alpha=\beta=1$), not much can be done and the rate is constant at $R_t=1/2$. The user has to query for both messages of $A$ and $B$ at every time $t$, \Ie $Q_t=AB$ for all $t$.

Figure~\ref{fig:th_plot} shows the rate $R_t$ as a function of time for different values of $\alpha+\beta$. As $\alpha+\beta$ approaches $1$, the correlation between the request decreases leading to an increase in the rate.

In the following section, we give the scheme that achieves the rate tuples given in Theorem \ref{thm:main}. 
We prove the converse in Appendix \ref{Subsec:Converse}.

\section{Achievability of Theorem \ref{thm:main}}
\label{Subsec:Achievable}

\subsection{ON-OFF Privacy Scheme}

In this section, we will describe an ON-OFF privacy scheme that achieves the rate in Theorem \ref{thm:main}, by specifying its encoding functions $\{\phi_t,\rho_t\}$ defined in Section~\ref{sec:formulation}. 

Our coding scheme retrieves of the messages in uncoded form. More specifically, the alphabet for the queries is $\cQ=~\{A,B,AB\}$.  The query values $A,B$ and $AB$ denote respectively  the user requesting the latest message of source $A$, $B$ or both. Upon receiving $Q_t \in \cQ$, the server responds by sending either one or two messages, such that
\begin{equation*}
  	Y_t = \rho_t \left(Q_t,W_{A,t},W_{B,t}\right) = 
  	\begin{cases}
  		W_{A,t}, & Q_t = A, \\
  		W_{B,t}, & Q_t = B, \\
  		\{W_{A,t},W_{B,t}\}, & Q_t = AB. 
  	\end{cases}
\end{equation*} 
The length of the answer $\ell\left(Q_t\right)$ is given by
\begin{equation*}
  	\ell\left(Q_t\right) = 
  	\begin{cases}
  		L, & Q_t = A~\text{or}~B, \\
  		2L, & Q_t = AB. 
  	\end{cases}
\end{equation*}
The normalized average length is
\begin{equation}
\label{eq:achievable-rate}
	\frac{\ell_t}{L} = 1+\Pr\left(Q_t=AB\right).
\end{equation} 

It remains to specify the query encoding functions $\{\phi_t\}$. The query encoding function $\phi_t$ at time $t$ is described as follows:
\begin{itemize}
	\item For $t \leq 0$, we simply download two messages to guarantee privacy, \Ie $Q_t=AB$. This is an immediate result in information-theoretic single-server private information retrieval  \cite{Chor_1995}. 
	\item For $t \geq 1$, the query $Q_t$ is a function of $Q_{t-1}$, $X_0$, $X_t$ and the local randomness $S_t$, \Ie 
	\[Q_t = \phi_t \left(X_0,X_t,Q_{t-1},S_t \right).\]
	Since we are not interested in the local randomness used, instead of specifying the function $\phi_t$ explicitly, we regard $Q_t$ as a probabilistic function of $\{X_0,X_t,Q_{t-1}\}$, and the distribution $p\left(Q_t|X_0,X_t,Q_{t-1} \right)$ is as follows:
	
	\vspace{10pt}
\begin{em}

Given $X_0$, $X_t$, and $Q_{t-1}$, 
\begin{enumerate}

\item if $Q_{t-1}\neq AB$, then $Q_t=X_t$ with probability $1$.
\item if $Q_{t-1}= AB$, then $p(Q_t\mid X_0,X_t, Q_{t-1})$ is as given in Table \ref{table:Qt}. 
\end{enumerate}
\end{em}
\end{itemize}
\begin{table*}[!htp]	
\normalsize
    
  \hspace{25pt}

      \centering
         \begin{tabular}{l|ccc||ccc||ccc}
	\diagbox[dir=SE,width=58pt,height=15pt]{\scriptsize$X_{0},X_{t}$}{\scriptsize$Q_{t}$} & $A$ & $B$ & $AB$ &$A$ & $B$ & $AB$&$A$ & $B$ & $AB$\\[\smallskipamount]
	\toprule
	$A,A$ &     $\frac{\beta}{1-\alpha}$            & 0     &  $\frac{1-\alpha - \beta}{1-\alpha}$&$\frac{1-\alpha}{\beta}$            & 0     &  $\frac{\alpha + \beta-1}{\beta}$&1&0&0\\[\smallskipamount]
	
	$A,B$ & 0  &      $1$              &     0 &$ 0$  &      $1$              &     $0$    &0&    $\frac{1-\beta}{\alpha}$              &     $\frac{\alpha+\beta-1}{\alpha}$    \\[\smallskipamount]
	
	$B,A$ &       $1$          & 0     &  0    &$1$  &  0&0  &  $ \frac{1-\alpha}{\beta}$    &0        &  $\frac{\alpha + \beta-1}{\beta}$ \\[\smallskipamount]
	
	$B,B$ & 0  &  $ \frac{\alpha}{1-\beta}$                    &  $\frac{1-\alpha - \beta}{1-\beta}$&0  &  $ \frac{1-\beta}{\alpha}$                    &  $\frac{\alpha + \beta-1}{\alpha}$&0  &  1  & 0\\[\smallskipamount]
	\toprule
	\multicolumn{1}{c}{} & \multicolumn{3}{c}{(a) $\alpha+\beta< 1$} & \multicolumn{3}{c}{(b) $\alpha+\beta> 1$ and $t$ is even} & \multicolumn{3}{c}{(c) $\alpha+\beta> 1$ and $t$ is odd} 
	\end{tabular}
	  \caption{ The proposed   ON-OFF privacy scheme achieving capacity. The query $Q_t$ is probabilistic and depends on the current request $X_t$, the previous query $Q_{t-1}$ and the last private request $X_0$.  If $Q_{t-1}\neq AB$ then  $Q_t=X_t$. Otherwise, $Q_{t-1}$ is chosen based on  the probabilities $p(Q_t\mid X_0,X_t, Q_{t-1}=AB)$ given in this table for (a) $\alpha+\beta<1$, (b) and (c) are for $\alpha+\beta>1$ where $t$ is even or odd respectively. }
	   \label{table:Qt}
	   \vspace{-5pt}
\end{table*}
\label{sec:main_results}
%
%


\subsection{Privacy}
In this subsection, we prove that the given scheme satisfies the privacy constraint for $t \geq 1$. 
Recall the privacy constraint \eqref{eq:privacy} that $I \left(X_{\cB_t};Q_t|Q_{(t-1)}\right) = 0$, where $\cB_t = \{i: i \leq 0, i \in \bbZ\}$. 
We want to show that
\vspace{-5pt}
\begin{align}
   &I \left(X_{\cB_t};Q_t|Q_{(t-1)}\right) \nonumber\\
   &= I \left(X_{0};Q_t| Q_{(t-1)} \right)+ I \left(X_{\cB_t\backslash\{0\}};Q_t|X_{0}, Q_{(t-1)}\right) \label{eqn:privacy_main}\\
   &=0,\nonumber
\end{align}
To do that we will show that each of the terms in the sum in (\ref{eqn:privacy_main}) is equal to zero.

\begin{claim}
$I \left(X_{\cB_t\backslash\{0\}};Q_t|X_{0}, Q_{(t-1)}\right) = 0$.
\label{claim:privacy_main_1}
\end{claim}
The claim can be justified as follows:
\begin{align*}
  & I \left(X_{\cB_t\backslash\{0\}};Q_t|X_{0}, Q_{(t-1)}\right) \\
  & = H \left(X_{\cB_t\backslash\{0\}}|X_{0}, Q_{(t-1)}\right)  - H \left(X_{\cB_t\backslash\{0\}}|X_{0}, Q_{(t)}\right) \\
  & \leq H \left(X_{\cB_t\backslash\{0\}}|X_{0}, Q_{\cB_t}\right) - H \left(X_{\cB_t\backslash\{0\}}|X_{0}, Q_{(t)}\right) \\
  & = H \left(X_{\cB_t\backslash\{0\}}|X_{0}, Q_{\cB_t}\right) - H \left(X_{\cB_t\backslash\{0\}}|X_{0}, Q_{\cB_t}, Q_{[1:t]}\right) \\
  & \utag{a}{\leq} H \left(X_{\cB_t\backslash\{0\}}|X_{0}, Q_{\cB_t}\right) \hspace{-2pt}- \hspace{-2pt}H \left(X_{\cB_t\backslash\{0\}}|X_{0}, Q_{\cB_t}, X_{[1:t]},S_{[1:t]}\right) \\
  & \utag{b}{=} H \left(X_{\cB_t\backslash\{0\}}|X_{0}, Q_{\cB_t}\right) - H \left(X_{\cB_t\backslash\{0\}}|X_{0}, Q_{\cB_t}\right) \\
  &  = 0,
\end{align*}
where \uref{a} follows because $Q_{[1:t]}$ is a function of $\left\{X_{[0:t]},S_{[1:t]}\right\}$, and \uref{b} follows from the independence between $\{X_i: i \in \bbZ\}$ and $\{S_i:i \in \bbZ\}$, and the Markovity of $\{X_i: i \in \bbZ\}$. 


\begin{claim}
\label{claim:privacy_main_2}
$
	I \left(X_{0};Q_t|Q_{(t-1)}\right) = 0
$
for $t \geq 1$.
\end{claim}
The proof of Claim \ref{claim:privacy_main_2} can be found in Appendix \ref{app:proof_lemma_private}.
  


\subsection{Rate}
Now, we evaluate the rate achieved by this coding scheme. We know from \eqref{eq:achievable-rate} that 
\[\frac{1}{R_t} = 1 + \Pr\left(Q_t=AB\right) \]
is achievable. For $t \leq 0$, since $\Pr\left(Q_t=AB\right) = 1$, we know that $R_t = \frac{1}{2}$ is achievable. To complete the computation of the rate, for $t \geq 1$, we need the following result in Lemma \ref{Lem:Q} whose proof can be found in Appendix \ref{app:proof_lemma_Q}.
\begin{lemma}
\label{Lem:Q}
The random variables $\{Q_t: t \geq 0\}$ form a Markov chain with transition matrix $P$, where
\begin{equation}
\label{eq:transition-P1}
  P = \begin{bmatrix}
        1- \alpha &  \alpha  &    0  \\
        \beta &  1- \beta & 0 \\
       \beta & \alpha & 1- \alpha -\beta
\end{bmatrix}, \quad \text{ if } \alpha+\beta\leq 1,
\end{equation}
and 
\begin{equation}
\label{eq:transition-P2}
  P = 
\begin{bmatrix}
        1- \alpha &  \alpha  &    0  \\
        \beta &  1- \beta & 0 \\
       1- \alpha & 1- \beta & \alpha +\beta-1
\end{bmatrix},\quad \text{ if } \alpha+\beta> 1.
\end{equation}
\end{lemma}

From Lemma~\ref{Lem:Q}, we easily obtain that 
 \begin{small}
\begin{align*}
	 \Pr\left(Q_t=AB \right) & \utag{a}{=} \Pr\left(Q_t=Q_{t-1}=\cdots=Q_0=AB \right) \\
	& = \Pr\left(Q_0=AB \right)\prod_{i=1}^t \Pr\left(Q_i=AB|Q_{i-1}=AB \right) \\
	& \utag{b}{=} \prod_{i=1}^t \Pr\left(Q_i=AB|Q_{i-1}=AB \right),
\end{align*}
 \end{small}
where \uref{a} follows because
\[\Pr\left(Q_i=AB|Q_{i-1}=A \right) = \Pr\left(Q_i=AB|Q_{i-1}=B \right) = 0\]
for the transition matrices given in both \eqref{eq:transition-P1} and \eqref{eq:transition-P2}; and \uref{b} follows from $\Pr\left(Q_0=AB \right) =1$, which can be justified because the user is required to download both messages at $t=0$ since $F_0=\text{ON}$. 

Using 
\eqref{eq:transition-P1} and \eqref{eq:transition-P2}, we have
\begin{equation*}
  \Pr\left(Q_t=AB \right) = |1- \alpha -\beta|^t.
\end{equation*}
Therefore, we can conclude that 
  \begin{equation*}
      R_t \leq \frac{1}{1+ |1- \alpha - \beta|^t}.
  \end{equation*}
 is achievable  for $t \geq 1$.

\section*{Acknowledgment}
This work was supported by NSF Grant CCF 1817635.


\appendices
\section{Converse of Theorem~\ref{thm:main}}
\label{Subsec:Converse}

In this section, we will prove the converse. For $t \leq~0$, we know from \cite{Chor_1995} that it is necessary to download two messages to achieve perfect privacy. For $t \geq~1$, we will show that for any given $\{\phi_t,\rho_t\}$ satisfying the decodable condition and the privacy constraint, the rate is upper bounded by
\[R_t \leq \frac{1}{1+ |1- \alpha -\beta|^t},\]
or equivalently
\[\ell_t/L \geq 1+ |1- \alpha -\beta|^t.\]
Since 
\[\ell_t = \mathbb{E}_{Q_t}[\ell(Q_t)] = \sum_{q \in \cQ}\Pr\left(Q_t=q\right)\ell(q),\]
we consider partitioning the alphabet $\cQ$  into three disjoint subsets $\cQ_a$, $\cQ_b$ and $\cQ_{ab}$ based on the decodability of $W_{A,t}$, $W_{B,t}$ or $\{W_{A,t},W_{B,t}\}$. Roughly speaking,  $\rho_t\left(q \in \cQ_a, W_{A,t},W_{B,t} \right)$ can decode $W_{A,t}$ correctly but cannot decode $W_{B,t}$. Similarly, $\rho_t\left(q \in \cQ_b, W_{A,t},W_{B,t} \right)$ can decode $W_{B,t}$ correctly but cannot decode $W_{A,t}$, and $\rho_t\left(q \in \cQ_{ab}, W_{A,t},W_{B,t} \right)$ can decode both $W_{A,t}$ and $W_{B,t}$ correctly. Clearly, $\ell (q \in \cQ_a) \geq L$, $\ell (q \in \cQ_b) \geq L$ and $\ell (q \in \cQ_{ab}) \geq 2L$. Hence, we have
\begin{equation}
\label{eq:converse-length-probability}
\ell_t/L \geq 2 - \Pr\left(\cQ_a \cup \cQ_b \right).  
\end{equation}

Recall the privacy constraint for $i \geq 1$, 
\begin{equation}
\label{eq:converse-pricacy-t1}
  I \left(X_{\cA_0};Q_i|Q_{(i-1)}\right) = 0,
\end{equation} 
where $\cA_0=\{\ldots,-2,-1,0\}$.
Since \eqref{eq:converse-pricacy-t1} holds for any $i \geq 1$, for a fixed $t$, we have
\begin{align*}
  I \left(X_{\cA_0};Q_{(t)}\right) 
  & = I \left(X_{\cA_0};Q_{[\cA_0]}\right) + \sum_{i=1}^t I \left(X_{\cA_0};Q_{i}|Q_{(i-1)}\right)\\
  & = I \left(X_{\cA_0};Q_{[\cA_0]}\right) \\
  & = 0. 
\end{align*}
From $I \left(X_{\cA_0};Q_{(t)}\right)  = 0$, we can easily have
\begin{equation}
\label{eq:converse-privacy}
  I \left(X_0; Q_{t} \right) = 0,
\end{equation}
and \eqref{eq:converse-privacy} can be written as 
\begin{equation}
\label{eq:privacy-Q-probability}
\Pr\left(Q_t=q|X_0 = A\right) = \Pr\left(Q_t=q|X_0 = B\right),~ \forall q \in \cQ.
\end{equation}

Now, we focus on the marginal distribution $p\left(Q_t,X_0,X_t\right)$. For notational simplicity, let $P(A,A)=\Pr\left(X_0=A,X_t=A\right)$ and $P(A|A)=\Pr\left(X_t=A|X_0=A\right)$. Here, $P(A,B)$, $P(B,A)$, $P(B,B)$ and $P(A|B)$, $P(B|A)$, $P(B|B)$ are defined similarly. Also, let $\delta= \Pr\left(X_0=A\right)$ and $1 - \delta= \Pr\left(X_0=B\right)$. 

By referring to the decodability and \eqref{eq:privacy-Q-probability}, we know that any adimissible $p\left(Q_t,X_0,X_t\right)$ can be illustrated by Table~\ref{Tab:converse}. 

\begin{table}[h]
\normalsize
  \centering
  \begin{tabular}{l|ccc}
  $(X_{0},X_{t})$ & $Q_t\in\cQ_a$ & $Q_t\in\cQ_b$ & $Q_t\in\cQ_{ab}$ \\

  \toprule
  $(A,A)$ &     $p_1$            & 0     &  $P(A,A) - p_1$                 \\   [\medskipamount]
  $(A,B)$ &     0          &  $p_2$    &    $P(A,B) - p_2$    \\[\medskipamount]
  
  $(B,A)$ &     $\frac{1- \delta}{\delta} p_1$          & 0     &    $P(B,A) - \frac{1- \delta}{\delta} p_1$          \\[\medskipamount]
  
  $(B,B)$ & 0  &      $\frac{1- \delta}{\delta} p_2$               &   $P(B,B)- \frac{1- \delta}{\delta} p_2$           \\
  
  \end{tabular}
  \caption{The joint distribution  $p(Q_t,X_0,X_t)$ \\ }
  \label{Tab:converse}
\end{table}

By examining the values in the table, we have
\begin{equation*}
  \begin{aligned}
    0 & \leq \frac{p_1}{\delta} & \leq \min\left\{P(A|A), P(A|B)\right\},  \\
    0 & \leq \frac{p_2}{\delta} & \leq \min\left\{P(B|A), P(B|B)\right\}.
  \end{aligned}   
\end{equation*} 
Hence, we obtain that
\begin{align*}
  \ell_t/L & \geq 2 - \Pr\left(\cQ_a \cup \cQ_b \right) \\
  & = 2 - \frac{p_1 + p_2}{\delta} \\
  & \geq  2 - \min\left\{P(A|A), P(A|B)\right\} \\
  &\hspace{110pt}- \min\left\{P(B|A), P(B|B)\right\}.
\end{align*}
From the Markovity of $\{X_t:t \in \bbZ\}$, we have
\[
\small
\begin{aligned}
  P(A|A)  = \frac{\beta+\alpha(1- \alpha -\beta)^t}{\alpha+\beta},~ & P(B|A)  = \frac{\alpha-\alpha(1- \alpha -\beta)^t}{\alpha+\beta},\\
  P(A|B)  = \frac{\beta- \beta(1- \alpha -\beta)^t}{\alpha+\beta},~ & P(B|B)  = \frac{\alpha+\beta(1- \alpha -\beta)^t}{\alpha+\beta}.\\
\end{aligned}
\]
Therefore, we finally obtain that 
\[\ell_t/L \geq  1+ |1- \alpha -\beta|^t,
\]
which completes the converse proof.

\section{Proof of Claim \ref{claim:privacy_main_2}}
\label{app:proof_lemma_private}
We first introduce three propositions. They show the dependency relations between random variables induced by the given coding scheme. The propositions are straightforward so the proofs are omitted. 
\begin{proposition}
\label{Prop:X}
  For $t \geq 0$, $X_t$ is a deterministic function of $Q_t$ and $X_0$, \Ie $X_{t}=g(X_0,Q_t)$.
\end{proposition}
\begin{proposition}
\label{Prop:Markov-X}
  For $t \geq 1$,  $\{X^{(t-1)},Q^{(t-1)}\}\rightarrow X_{t-1}\rightarrow X_t$ forms a Markov chain. In particular, any subset of $\{X^{(t-1)},Q^{(t-1)}\}$ is independent of $X_t$ given $X_{t-1}$.
\end{proposition}
\begin{proposition}
\label{Prop:Markov-Code}
  For $t \geq 1$, $\{X^{(t-1)},Q^{(t-1)}\}\rightarrow \{X_t,X_0,Q_{t-1} \} \rightarrow Q_t$ forms a Markov chain. In particular, any subset of $\{X^{(t-1)},Q^{(t-1)}\}$ is independent of $Q_t$ given $\{X_t,X_0,Q_{t-1} \}$.
\end{proposition}
\noindent
Claim \ref{claim:privacy_main_2} is equivalent to
\begin{multline*}
   \Pr\left(Q_t=q|X_0=A, Q_{[t-1]}=\mathbf{\bar{q}}\right) \\
   = \Pr\left(Q_t=q|X_0=B, Q_{[t-1]}=\mathbf{\bar{q}}\right),
\end{multline*} 
for any $q \in \cQ$ and $\mathbf{\bar{q}} \in {\displaystyle\prod_{i \leq t-1}}\cQ$. Therefore consider,
\begin{small}
\begin{align*}
   &\hspace{-55pt}\Pr\left(Q_t=q|X_0=x_0, Q_{(t-1)}=\mathbf{\bar{q}}\right)  \\
   &\hspace{-55pt}= \Pr\left(Q_t=q|X_0=x_0, Q_{t-1}=q', Q_{(t-2)}= \mathbf{\bar{q}'}\right) 
  \end{align*}
  \vspace{-20pt}
   \begin{align*}
   = \sum_{x}  &\Pr\left(X_t=x|X_0=x_0, Q_{t-1}=q', Q_{(t-2)}= \mathbf{\bar{q}'}\right)\times\\ 
   &\Pr\left(Q_t=q| X_t=x,X_0=x_0, Q_{t-1}=q', Q_{(t-2)}= \mathbf{\bar{q}'}\right) \\
   \utag{a}{=} \sum_{x} & \Pr\left(X_t=x|X_0=x_0, Q_{t-1}=q', Q_{(t-2)}= \mathbf{\bar{q}'}\right)\times\\ &\Pr\left(Q_t=q| X_t=x,X_0=x_0, Q_{t-1}=q'\right) \\
   \utag{b}{=} \sum_{x}&  \Pr\left(X_t=x|X_{t-1}\hspace{-3pt}=\hspace{-3pt}g(x_0,q'), X_0\hspace{-3pt}=\hspace{-3pt}x_0, Q_{t-1\hspace{-2pt}}=\hspace{-2pt}q', Q_{(t-2)}\hspace{-2.5pt}=\hspace{-2.5pt}\mathbf{\bar{q}'}\right) \\
  &\Pr\left(Q_t=q| X_t=x,X_0=x_0, Q_{t-1}=q'\right) \\
   \utag{c}{=} \sum_{x}&  \Pr\left(X_t=x|X_{t-1}=g(x_0,q')\right)\times \\
  &\Pr\left(Q_t=q| X_t=x,X_0=x_0, Q_{t-1}=q'\right),
\end{align*}
\end{small}where \uref{a} follows from Proposition~\ref{Prop:Markov-Code}, \uref{b} follows from Proposition~\ref{Prop:X}, and \uref{c} follows from Proposition~\ref{Prop:Markov-X} and the Markovity of  $\{X_i: i \in \bbZ\}$.

If $q'=A$ or $B$, we have
\begin{small}
    \begin{align}
    &\sum\limits_{x}\Pr\left(X_t=x| X_{t-1}=g(x_0,q')\right)\nonumber\\
    &\hspace{80pt}\Pr\left(Q_t=q| X_t=x,X_0=x_0, Q_{t-1}=q'\right)\nonumber\\
    &\utag{a}{=} \sum_{x}  \Pr\left(X_t=x|X_{t-1}=q'\right)\nonumber\\ &\hspace{80pt}\Pr\left(Q_t=q| X_t=x,X_0=x_0, Q_{t-1}=q'\right) \nonumber\\
     & \utag{b}{=} \sum_{x}  \Pr\left(X_t=x|X_{t-1}=q'\right) \Pr\left(Q_t=q| X_t=x, Q_{t-1}=q'\right),\label{eq:proof-temple} 
    \end{align} 
\end{small}where \uref{a} follows from the fact that if $Q_{t-1}=A$ or $B$ then $X_{t-1}=Q_{t-1}$, and \uref{b} follows because given $Q_{t-1}=A$ or $B$, $Q_t=X_t$ with probability $1$. 

Clearly, R.H.S of \eqref{eq:proof-temple} is independent of the choice of $x_0$, and thus it remains to show that 
\begin{small}
\begin{align}
   &\sum_{x=A,B}  \Pr\left(X_t=x|X_{t-1}=g(A,AB)\right)\nonumber\\ &\hspace{80pt}\Pr\left(Q_t=q| X_t=x,X_0=A, Q_{t-1}=AB\right) \nonumber\\
   &= \sum_{x=A,B}  \Pr\left(X_t=x|X_{t-1}=g(B,AB)\right)\nonumber \\
   &\hspace{80pt}\Pr\left(Q_t=q| X_t=x,X_0=B, Q_{t-1}=AB\right),
   \label{eq:proof-lemma-1} 
\end{align}
\end{small}for any $q \in \{A,B,AB\}$. Towards this end, let us discuss separately as follows:
\begin{itemize}
  \item When $\alpha+\beta \leq 1$, we have
  \begin{small}
  \begin{align}
     &\sum_{x=A,B}  \Pr\left(X_t=x|X_{t-1}=g(x_0,AB)\right)\nonumber\\ &\hspace{60pt}\Pr\left(Q_t=q| X_t=x,X_0=x_0, Q_{t-1}=AB\right) \nonumber\\
     &\utag{a}{=} \sum_{x=A,B} \Pr\left(X_t=x|X_{t-1}=x_0\right)\nonumber\\ &\hspace{60pt}\Pr\left(Q_t=q|X_t=x,X_0=x_0, Q_{t-1}=AB\right),
     \label{eq:proof-lemma-2} 
  \end{align}
  \end{small}where \uref{a} follows Table~\ref{table:Qt}(a) where $X_{t-1}=X_0$ given $Q_{t-1}=AB$.

  Substituting $x_0$ by $A$ and $B$ in  \eqref{eq:proof-lemma-1} on the L.H.S and R.H.S. respectively, we can verify from Table \ref{table:Qt}(a) and transition matrix $M$ that
  \begin{small}
  \begin{align}
   &\sum_{x=A,B}    \Pr\left(X_t=x|X_{t-1}=A\right)\nonumber\\ &\hspace{60pt}\Pr\left(Q_t=q|X_t=x,X_0=A, Q_{t-1}=AB\right) \nonumber\\
   &= \sum_{x=A,B}  \Pr\left(X_t=x|X_{t-1}=B\right)\nonumber\\ &\hspace{60pt}\Pr\left(Q_t=q|X_t=x,X_0=B, Q_{t-1}=AB\right)
   \label{eq:proof-lemma-3} 
\end{align}
\end{small}for all $q$. 

\item When $\alpha+\beta \leq 1$ and $t$ is odd, $t-1$ is even, and from Table \ref{table:Qt}(b), $Q_{t-1}=AB$ only if $X_{t-1}=X_0$. Therefore, \eqref{eq:proof-lemma-2} still holds, and we can verify \eqref{eq:proof-lemma-3} from Table \ref{table:Qt}(c) and the transition matrix $M$.

\item When $\alpha+\beta \leq 1$ and $t$ is even, $t-1$ is odd, and from Table \ref{table:Qt}(c), $Q_{t-1}=AB$ only if $X_{t-1}\neq  X_0$, and we have
\begin{small}
    \begin{align}
   &\sum_{x=A,B}    \Pr\left(X_t=x|X_{t-1}=B\right)\nonumber\\
   &\hspace{60pt}\Pr\left(Q_t=q|X_t=x,X_0=A, Q_{t-1}=AB\right) \nonumber\\
   &= \sum_{x=A,B}  \Pr\left(X_t=x|X_{t-1}=A\right) \nonumber\\
   &\hspace{60pt}\Pr\left(Q_t=q|X_t=x,X_0=B, Q_{t-1}=AB\right)
   \label{eq:proof-lemma-4} 
\end{align}
\end{small}for all $q$. Similarly, we can verify (\ref{eq:proof-lemma-4}) from Table \ref{table:Qt}(b) and the transition matrix $M$.
   
\end{itemize}





\section{Proof of Lemma~\ref{Lem:Q}}
\label{app:proof_lemma_Q}
First, $Q_t=A$ or $B$ only if $Q_{t-1}\neq AB$; therefore it is easy to see the following,
\begin{align*}
&\Pr\left(Q_t=A|Q_{t-1}=A\right)= \Pr\left(X_t=A|X_{t-1}=A\right)=1- \alpha,&\\
&\Pr\left(Q_t=B|Q_{t-1}=A\right)= \Pr\left(X_t=B|X_{t-1}=A\right)= \alpha,&\\
&\Pr\left(Q_t=A|Q_{t-1}=B\right)= \Pr\left(X_t=A|X_{t-1}=B\right)= \beta,&\\
&\Pr\left(Q_t=B|Q_{t-1}=B\right)=
\Pr\left(X_t=B|X_{t-1}=B\right)= 1- \beta.&
\end{align*}
Then, we consider 
\begin{small}
\begin{align*}
  & \Pr\left(Q_t=q|Q_{t-1}=AB\right) \\
  & = \sum_{x_0,x_{t-1},x_t} \hspace{-10pt}\Pr\left(Q_t\hspace{-2pt}=\hspace{-2pt}q,X_0=x_0,X_{t-1}=x_{t-1}\hspace{-2pt},X_t\hspace{-2pt}=\hspace{-2pt}x_t|Q_{t-1}\hspace{-2pt}=\hspace{-2pt}AB\right) \\
  & = \sum_{x_0,x_{t-1},x_t}  \hspace{-10pt}\Pr\left(X_0=x_0,X_{t-1}=x_{t-1}|Q_{t-1}=AB\right) \times\\&\hspace{25pt}\Pr\left(Q_t=q,X_t=x_t|X_0=x_0,X_{t-1}=x_{t-1},Q_{t-1}=AB\right)  \\
  & \utag{a}{=} \hspace{-18pt} \sum_{x_0,x_{t-1}=g(x_0,AB),x_t}  \hspace{-30pt}\Pr\left(X_0=x_0|Q_{t-1}=AB\right) \times\\&\hspace{25pt}\Pr\left(Q_t=q,X_t=x_t|X_0=x_0,X_{t-1}=x_{t-1},Q_{t-1}=AB\right) \\
  & \utag{b}{=} \hspace{-18pt}\sum_{x_0,x_{t-1}=g(x_0,AB),x_t}\hspace{-30pt} \Pr\left(X_0=x_0\right) \times\\&\hspace{25pt}\Pr\left(Q_t=q,X_t=x_t|X_0=x_0,X_{t-1}=x_{t-1},Q_{t-1}=AB\right) \\
  &  = \hspace{-18pt}\sum_{x_0,x_{t-1}=g(x_0,AB),x_t}\hspace{-30pt} \Pr\left(X_0=x_0\right) \times\\&\hspace{25pt}\Pr\left(X_t=x_t|X_0=x_0,X_{t-1}=x_{t-1},Q_{t-1}=AB\right)
  \times\\&\hspace{25pt}\Pr\left(Q_t=q|X_t=x_t,X_0=x_0,X_{t-1}=x_{t-1},Q_{t-1}=AB\right) \\
  & \utag{c}{=} \hspace{-18pt}\sum_{x_0,x_{t-1}=g(x_0,AB),x_t} 
  \hspace{-30pt}\Pr\left(X_0=x_0\right) \Pr\left(X_t=x_t|X_{t-1}=x_{t-1}\right) \times\\&\hspace{75pt}\Pr\left(Q_t=q|X_t=x_t,X_0=x_0,Q_{t-1}=AB\right),
\end{align*}
\end{small}where \uref{a} follows from Preposition \ref{Prop:X} where $X_{t-1}$ is a function of $X_0$ given $Q_{t-1}$, \uref{b} follows from the privacy at time $t-1$. The second term in \uref{c} follows from $Q_{t-1}$ being a function of $X_0$ and $X_{t-1}$ and the Markovity of  $\{X_i: i \in \bbZ\}$, and the third term follows from Proposition \ref{Prop:X}.

Now we substitute $q$ by $A$, $B$ and $AB$ and discuss two cases $\alpha+\beta \leq 1$ and $\alpha + \beta >1$.

\begin{itemize}
\item For $\alpha+\beta \leq 1$, $X_{t-1}=g(x_0,AB)=x_0$. Then, 
\begin{small}
\begin{multline*} 
\hspace{-12pt}\Pr\left(Q_t=q|Q_{t-1}=AB\right)=\\
\hspace{-15pt}\sum_{x_0,x_t} \Pr\left(X_0=x_0\right) \Pr\left(X_t=x_t\mid X_{t-1}=x_0\right)\times\\
 \Pr\left(Q_t=q|X_t=x_t,X_0=x_0,Q_{t-1}=AB\right).
 \end{multline*}
 \end{small}For instance, let $\Pr(X_0=A)=p_0$ and $\Pr(X_0=B)=1-p_0$, and using the values given in Table \ref{table:Qt}(a) and transition matrix $M$, we can verify that
 \begin{small}
 
 \begin{align*}
 \Pr\left(Q_t=A|Q_{t-1}=AB\right)&=p_0(1-\alpha)\frac{\beta}{1-\alpha}+(1-p_0)\beta\\&=\beta.
 \end{align*}
 
 \end{small}Similarly, we can verify the rest of values given in transition matrix $P$ for $\alpha+\beta\leq 1$.
 
 \item For $\alpha+\beta > 1$, 
 \begin{itemize}
 \item[\scriptsize{$\bullet$}] if $t$ is odd, then $t-1$ is even, and
 $$X_{t-1}=g(x_0,AB)=
 \begin{cases}
 A&x_0=B,\\
 B&x_0=A.
 \end{cases}$$
  \item[\scriptsize{$\bullet$}] if $t$ is even, then $t-1$ is odd, and
 $$X_{t-1}=g(x_0,AB)=x_0.$$
 \end{itemize}
We can verify the remaining elements of the transition matrix $P$, for $\alpha+\beta>1$, using the values in transition matrix $M$, and the values in Table \ref{table:Qt}(c) and (b), for $t$  odd and even respectively.
\end{itemize}

\section{General Privacy Mode}
\label{app:discussion}
So far, we have focused on the privacy mode being the step function described in \eqref{eq:step-function}. When the privacy mode is an arbitrary sequence, we can generalize the result of Theorem~\ref{thm:main}.
So the rate tuple $\left(R_t:t \in \bbZ\right)$ is achievable if and only if  
  \begin{equation}
      R_t \leq 
      \begin{cases}
      \frac{1}{2}, & F_t = \text{ON},\\
        \frac{1}{1+ |1- \alpha -\beta|^{t-F^{-}(t)}}, & F_t = \text{OFF},
      \end{cases}
      \label{eq:rate_general}
  \end{equation}
where $F^{-}(t) = \max\{i: i \leq t, F_i = \text{ON}\}$, \Ie $F^{-}(t)$ is the latest time the privacy was ON. 

The intuition is the following. To protect all the past requests when privacy was ON, it suffices to protect the last request when privacy was ON, which is $F^-(t)$. This follows mainly from the Markovity of the requests.

The proof of \eqref{eq:rate_general} when $F_t=\text{OFF}$ follows similar steps as the proof of Theorem \ref{thm:main}. In particular, in the converse proof,
by applying the chain rule to $\sum_{i=F^{-}(t)+1}^t \left(X_{\cB_i};Q_{i}|Q_{(i-1)}\right)$, we can easily obtain that
\[I \left(X_{F^{-}(t)}; Q_{t}|Q_{(F^{-}(t))}\right) = 0.\]
Moreover,  instead of inspecting the distribution $p\left(Q_t,X_t,X_{(F^{-}(t))}\right)$ for the step function, we can inspect the distribution $p\left(Q_t,X_t,X_{(F^{-}(t))},Q_{(F^{-}(t))}\right)$ here. Note that for any fixed $q_{(F^{-}(t))}$, we have exactly the same proof as we did for the step function. Hence, we can  obtain the same upper bound on the rate, \Ie
  \begin{equation*}
      R_t \leq 
        \frac{1}{1+ |1- \alpha -\beta|^{t-F^{-}(t)}}.
  \end{equation*}

For the achievability proof when $F_t=\text{ON}$, the user downloads the messages from both sources. When $F_t=\text{OFF}$, the coding scheme is similar to before and can be obtained by replacing $X_0$ by $X_{F^{-}(t)}$, that is
\[Q_t = \phi_t \left(X_{F^{-}(t)},X_t,Q_{t-1},S_t\right).\]
Then, one can check that the obtained coding scheme satisfies the privacy constraint for any privacy mode $\{F_t: t \in \mathbb{Z}\}$. 
Moreover, it achieves the rate in \eqref{eq:rate_general}. The verification details are similar to those for the step function.

\end{document}

%% file: preamble.tex
\usepackage{graphicx}                                      
\usepackage{amssymb,amsthm,bm}  
\usepackage[cmex10]{amsmath} 

\usepackage{multirow}   
\usepackage{tikz}                   
\newtheorem{theorem}{Theorem}
\newtheorem{lemma}{Lemma}

\newtheorem{definition}{Definition}

\newtheorem{proposition}{Proposition}

\newtheorem{claim}{Claim}

\newcommand{\utag}[2]{\mathop{#2}\limits^{\text{(#1)}}}
\newcommand{\uref}[1]{(#1)}

\long\def\symbolfootnote[#1]#2{\begingroup
\def\thefootnote{\fnsymbol{footnote}}\footnote[#1]{#2}\endgroup}

\usepackage{xcolor}
\definecolor{SkyBlue}{RGB}{240,255,255}
\definecolor{LavenderBlush}{RGB}{218,112,214}
\definecolor{BlanchedAlmond}{RGB}{255,235,205}

\newcommand{\cA}{\mathcal{A}}
\newcommand{\cB}{\mathcal{B}}

\newcommand{\cN}{\mathcal{N}}

\newcommand{\cP}{\mathcal{P}}
\newcommand{\cQ}{\mathcal{Q}}

\usepackage{mathrsfs}

\newcommand{\sC}{\mathscr{C}}

\newcommand{\bbZ}{\mathbb{Z}}

\newcommand{\Ie}{\textit{i.e., }}

\usepackage{diagbox}
\usepackage{color}

\usepackage{tikz}
\usetikzlibrary{automata,positioning,chains,fit,shapes,calc,arrows,plotmarks,arrows.meta}
\usepackage{tkz-graph}
\usepackage{caption}
\usepackage{booktabs}
\usepackage{cite}
\usepackage{hyperref}

\makeatletter
\newcommand{\mathleft}{\@fleqntrue\@mathmargin0pt}
\newcommand{\mathcenter}{\@fleqnfalse}
\makeatother

\usepackage{pgfplots}
  \pgfplotsset{compat=newest}

  \usepgfplotslibrary{patchplots}
  \usepackage{grffile}

%% file: plot.tex
%
%
\definecolor{mycolor1}{rgb}{0.00000,0.44700,0.74100}%
\definecolor{mycolor2}{rgb}{0.85000,0.32500,0.09800}%
\definecolor{mycolor3}{rgb}{0.92900,0.69400,0.12500}%
\definecolor{mycolor4}{rgb}{0.49400,0.18400,0.55600}%
\begin{tikzpicture}

\begin{axis}[%
width=2.5in,
height=2in,
at={(1.011in,0.642in)},
scale only axis,
xmin=0,
xmax=20,
xtick={ 0,  2,  4,  6,  8, 10, 12, 14, 16, 18, 20},
xlabel style={font=\color{white!15!black}},
xlabel={Time $t$},
ymin=0.5,
ymax=1,
ylabel style={font=\color{white!15!black}},
ylabel={Download Rate $R_t$},
axis background/.style={fill=white},
legend style={at={(0.77,0.4)}, anchor=south west, legend cell align=left, align=left, draw=white!15!black}
]

\addlegendimage{empty legend}
\addlegendentry{\hspace{-.55cm}\textbf{$\alpha+\beta$}}

 \addlegendimage{empty legend}
 \addlegendentry{\hspace{-1.5cm} 
 }

\addplot [color=mycolor1, line width=1.5pt, mark size=2pt, mark=triangle, mark options={solid, rotate=90, mycolor1}]
  table[row sep=crcr]{%
0	0.5\\
1	1\\
2	1\\
3	1\\
4	1\\
5	1\\
6	1\\
7	1\\
8	1\\
9	1\\
10	1\\
11	1\\
12	1\\
13	1\\
14	1\\
15	1\\
16	1\\
17	1\\
18	1\\
19	1\\
20	1\\
};
\addlegendentry{1}

\addplot [color=mycolor2, line width=1.5pt, mark size=3pt, mark=x, mark options={solid, mycolor2}]
  table[row sep=crcr]{%
0	0.5\\
1	0.769230769230769\\
2	0.91743119266055\\
3	0.973709834469328\\
4	0.991965082829084\\
5	0.997575890585876\\
6	0.999271531053862\\
7	0.999781347819232\\
8	0.99993439430439\\
9	0.999980317387413\\
10	0.999994095134868\\
11	0.999998228533138\\
12	0.999999468559282\\
13	0.999999840567725\\
14	0.999999952170312\\
15	0.999999985651093\\
16	0.999999995695328\\
17	0.999999998708598\\
18	0.999999999612579\\
19	0.999999999883774\\
20	0.999999999965132\\
};
\addlegendentry{0.7}

\addplot [color=mycolor3, line width=1.5pt, mark size=2pt, mark=*, mark options={solid, mycolor3}]
  table[row sep=crcr]{%
0	0.5\\
1	0.625\\
2	0.735294117647059\\
3	0.822368421052631\\
4	0.885269121813031\\
5	0.927850356294537\\
6	0.955423749541397\\
7	0.972768702061958\\
8	0.98348129088135\\
9	0.990022850677816\\
10	0.993989724239196\\
11	0.996385144031034\\
12	0.997827945753317\\
13	0.998695634190672\\
14	0.999216971972409\\
15	0.999530035985447\\
16	0.99971796857342\\
17	0.999830762051884\\
18	0.999898450356709\\
19	0.999939067738966\\
20	0.9999634397523\\
};
\addlegendentry{0.4}

\addplot [color=mycolor4, line width=1.5pt, mark size=2pt, mark=diamond, mark options={solid, mycolor4}]
  table[row sep=crcr]{%
0	0.5\\
1	0.555555555555556\\
2	0.609756097560976\\
3	0.661375661375661\\
4	0.709421112372304\\
5	0.753193540612196\\
6	0.792302621570914\\
7	0.82664084901967\\
8	0.856331426842715\\
9	0.881664935499932\\
10	0.903037126829805\\
11	0.920895664738407\\
12	0.9356992379835\\
13	0.947889238046222\\
14	0.957872329434476\\
15	0.966011492215792\\
16	0.972623093734289\\
17	0.977977895569679\\
18	0.982304377486352\\
19	0.985793222434495\\
20	0.988602192725058\\
};
\addlegendentry{0.2}

\end{axis}
\draw[line width=0.4 pt] (7.48,5.6) -- (8.9,5.6);
\end{tikzpicture}%